\documentclass[aps,prl,twocolumn,superscriptaddress]{revtex4-2}
\usepackage{graphicx}
\usepackage{amssymb}
\usepackage{amsfonts}
\usepackage{amsmath}
\usepackage{natbib}
\usepackage{hyperref}
\usepackage{color}
\usepackage{bm}
\usepackage{txfonts}

\begin{document}

\title{Self-localized topological states in three dimensions}
\thanks{R.L. and P.L. contributed equally to this work.}

\author{Rujiang Li}
\email{rujiangli@xidian.edu.cn}
\affiliation{Key Laboratory of Antennas and Microwave Technology, School of
Electronic Engineering, Xidian University, Xi'an 710071, China}
\author{Pengfei Li}
\affiliation{Department of Physics, Taiyuan Normal University, Jinzhong 030619,
China} 
\affiliation{Institute of Computational and Applied Physics, Taiyuan Normal University,
Jinzhong 030619, China}
\author{Yongtao Jia}
\affiliation{Key Laboratory of Antennas and Microwave Technology, School of
Electronic Engineering, Xidian University, Xi'an 710071, China}
\author{Ying Liu}
\email{liuying@mail.xidian.edu.cn}
\affiliation{Key Laboratory of Antennas and Microwave Technology, School of
Electronic Engineering, Xidian University, Xi'an 710071, China}

\date{\today}

\begin{abstract}
Three-dimensional (3D) topological materials exhibit much
richer phenomena than their lower-dimensional counterparts. 
Here we propose self-localized topological states 
(i.e. topological solitons) in a 3D nonlinear photonic Chern insulator.
Despite being in the bulk and self-localized in all 3D, the topological solitons
at high-symmetry points $K$ and $K^{\prime}$ rotate in the same 
direction,
due to the underlying topology. Specifically, under the saturable nonlinearity
the solitons are stable over a broad frequency range. 
Our results highlight how topology and nonlinearity interact with 
each other and can be extended to other 3D topological systems.

\end{abstract}

\maketitle

\emph{Introduction}.---Since the discovery of the quantum Hall effect 
and its topological interpretation, extensive efforts have been put into the 
research of exotic topological materials \cite{RMP82-3045,RMP83-1057}. 
Dimensionality plays a key role in the classification of topological materials 
and determination of the topological states \cite{AIP-Conf,NJP12-065010,PRB90-165114}. 
Since for a realistic material three is the largest number of spatial dimensions 
in which electrons can move, 3D topological materials including
Weyl semimetals, 3D topological insulators, and 3D Chern insulators
gain particular attention \cite{RMP90-015001,NRP3-283}. 
In recent years, various engineered systems
have been implemented as the classical analogues of 3D topological
materials \cite{nphys12-337,nphoton11-130,nature565-622,ncommun11-2318,
nphys15-1150,nphys14-30,nphys11-920,science349-622,nphys13-611}.
Among them, the 3D photonic topological materials support robust
photonic propagation along a non-planar surface, which may find applications 
in topological lasers and photonic circuits \cite{nphys12-337,nphoton11-130,
nature565-622}. In these studies,
the interaction between photons is neglected.

In topological photonics, it is straightforward to include the interparticle 
interactions. Under high intensity, photons can effectively interact in a
nonlinear optical medium with a intensity-dependent refractive index. 
Several forms of nonlinear refractive indices such as Kerr nonlinearity, 
competing nonlinearity, and saturatable nonlinearity exist \cite{book-soliton}, and they 
provide a fertile ground to study the interplay between topology and nonlinearity.
Nonlinear topological photonics arises with many opportunities for 
fundamental discoveries and new functionalities for photonic devices \cite{APR7-021306}. 
However, the vast majority of research is carried out in lower-dimensions.
The studies of 3D photonic topological materials with nonlinearity
acts on all three spatial dimensions are rare.

In this Letter we propose self-localized topological states (i.e. topological solitons)
which are solely induced by the nonlinearity in a 3D photonic Chern insulator.
The 3D Chern insulator is realized by stacking 2D Chern insulators in the 
vertical direction \cite{NRP3-283}. In the linear regime (low optical intensity), 
the 3D Chern insulator supports the 2D surface states with chiral propagation along the
surfaces. While for the topological states that we discovered in
the nonlinear regime, they are self-localized in the bulk of the Chern 
insulator, rather than localized on the exterior or extended
in the vertical direction. 
Due to the same underlying topology that shared with the linear surface
states, the topological solitons reside in the linear bulk bandgap
and solitons at high-symmetry points $K$ and $K^{\prime}$ 
rotate in the same direction.
Specifically, under the saturable nonlinearity, the topological solitons
are dynamically stable for a wide frequency range.

Our topological solitons in 3D differ from the previously
reported solitons in lower-dimensional topological materials. 
First, our topological solitons are self-localized in the bulk. 
They are fundamentally different to the edge solitons which
are localized at the structure exterior or domain wall due to the 
bulk-boundary correspondence of their linear host lattices \cite{PRL117-143901,
PRL128-093901,PRX11-041507,ncommun9-3991,ncommun11-1902,PRA103-053507,
PRL123-254103}. 
Second, our topological solitons are also different to the bulk solitons
\cite{PRL111-243905,science368-856,LPR13-1900223,PRA98-013827,PRL118-023901,
arxiv1904.10312}.
In the linear regime, by stacking 2D Chern insulators into a 3D Chern 
insulator, the chiral edge states change into chiral surface states which
are extended in the stacking direction.
In the nonlinear regime, the introduction of another spatial dimension
usually leads to soliton stripes \cite{book-soliton}. Our topological
solitons are self-localized also in the vertical direction,
where interlayer coupling is delicately compensated by nonlinearity. 
The principle is similar to the balance between diffraction and
nonlinearity in the propagation direction of an edge soliton \cite{PRL117-143901,
PRX11-041507,ncommun11-1902,PRA103-053507}.
Such self-localization is important for constructing
diffraction-free topological states in 3D topological materials
and designing 3D topological photonic devices.

\emph{Hamiltonian}.---We start from a general Hamiltonian
\begin{eqnarray}
H_{L}&=&\left( \nu _{0}\delta k_{z}+\nu _{0}^{\prime }\delta
k_{z}^{2}\right) \sigma _{0}
+\nu_{x}\delta k_{x}\sigma_{1}
+\nu_{y}\delta k_{y}\sigma_{2}  \notag \\ 
&&+\left( \nu _{z}\delta k_{z}+\nu _{z}^{\prime }\delta
k_{z}^{2}\right) \sigma _{3}+m\sigma _{3},  
\label{Hamiltonian_1}
\end{eqnarray}
where $\sigma_{0}$ is identity matrix,
$\sigma_{i}$ ($i = 1, 2, 3$) are Pauli matrices, $m$ is the effective mass, 
$\nu_{x(y)}$ is the group velocity in $x(y)$ direction, and $\nu_{0,z}$ and 
$\nu_{0,z}^{\prime}$ are the group velocity and group velocity dispersion (GVD) in
$z$ direction, respectively. When $\nu_{0}^{\prime} =\nu_{z}^{\prime} =0$
and $m = 0$, this Hamiltonian reduces to the typical Weyl 
Hamiltonian \cite{RMP90-015001}. 
Based on the Weyl Hamiltonian, first we
include the GVD terms with $\delta k_{z}^{2}$ which are necessary 
to study the nonlinear
effect. Specifically, when $\nu_{0}=\nu_{z} = 0$ the second order contributions
need to be considered. The resulting Hamiltonian corresponds to a semi-Weyl
point with linear dispersions in $x$ and $y$ directions, and quadratic dispersion
in $z$ direction (similar to the semi-Dirac point or hybrid Dirac 
point \cite{PRL102-166803,PRL123-195503,PRB104-235121}). 
Then we introduce the mass
term $m$ which opens a bandgap at the nodal point. Usually, a mass term can
be created by breaking the time-reversal symmetry and/or inversion
symmetry \cite{ncommun11_1873}.

Transforming to position space \cite{LPR13-1900223},  the Hamiltonian is
\begin{eqnarray}
H_{L}&=&-i\sigma _{0}\left( \nu _{0}\partial _{z}-i\nu _{0}^{\prime
}\partial _{z}^{2}\right)
-i \nu _{x}\sigma _{1}\partial_{x}
-i \nu _{y}\sigma _{2}\partial_{y}  \notag\\
&&-i\sigma _{3}\left( \nu _{z}\partial _{z}-i\nu _{z}^{\prime
}\partial _{z}^{2}\right) +m\sigma _{3},  
\label{Hamiltonian_2}
\end{eqnarray}
with $i\partial_{t}\Psi=H_{L}\Psi$ and $\Psi = \left( \psi_{A}, \psi_{B}\right)^{T}$.
We can extend the system into the nonlinear regime by adding a general nonlinear
term $H_{NL}=N_{0}\left( \Psi\right)\sigma_{0}+ 
N_{z}\left( \Psi\right)\sigma_{3}$ with $N_{0,z}\in C\left( \mathbb{R}
\right)$ and $N_{0,z}\left(0\right)=0$
to the original Hamiltonian, and the second term $N_{z}\left(\Psi\right)$
is equivalent to a nonlinearity-induced mass.
The whole Hamiltonian is $H=H_{L}+H_{NL}$, and it
can be splitted into two parts $H=H_{\varparallel} +H_{z}$ with 
\begin{eqnarray}
H_{\varparallel}&=&-i \nu_{x}\sigma _{1}\partial _{x}
-i \nu_{y}\sigma _{2}\partial_{y}
+m\sigma _{3}+ N_{z}\left( \Psi\right)\sigma_{3}, \label{H_parallel}\\
H_{z}&=&-i\sigma _{0}\left( \nu _{0}\partial _{z}-i\nu _{0}^{\prime
}\partial _{z}^{2}\right)
-i\sigma _{3}\left( \nu _{z}\partial _{z}-i\nu _{z}^{\prime
}\partial _{z}^{2}\right) \notag\\
&&+N_{0}\left( \Psi\right)\sigma_{0}.
\label{H_z}
\end{eqnarray}
Using the Hamiltonian $H_{\varparallel}$, we get a generalized nonlinear
Dirac equation. In the special case where $N_{z}\left( \Psi\right)
=N_{z}\left(\Psi^{\dag} \sigma_{3} \Psi \right)$,
the Gross-Neveu/Soler type of nonlinear 
Dirac equation supports the Dirac solitons, which are topological
solitons in 2D \cite{PRL116-214101,
PRA98-013827,PRE100-022210}. The Hamiltonian $H_{z}$ also
admits the existence of solitons in $z$ direction, 
provided that the interlayer coupling governed
by $\partial _{z}^{2}$ is balanced with the nonlinear term $N_{0}\left( \Psi\right)$
\cite{book-soliton}. This principle has been used to realize
the edge solitons \cite{PRL117-143901,
PRX11-041507,ncommun11-1902,PRA103-053507}. Thus, the whole
Hamiltonian $H$ should support topological solitons that are
self-localized in all 3D.

\begin{figure}[tbp]
\includegraphics[width=8.08cm]{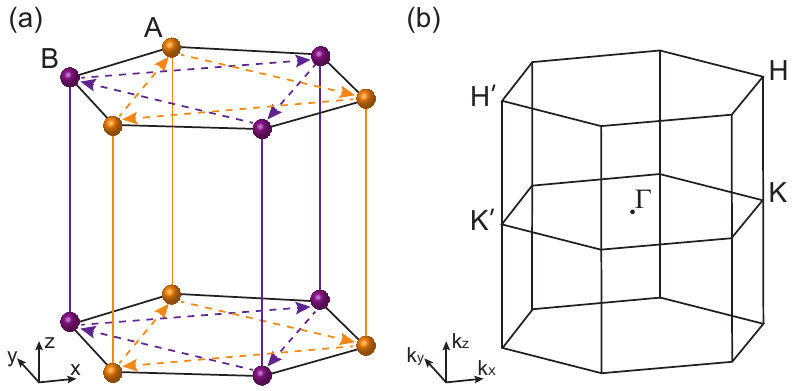}
\caption{(a) A 3D Chern insulator
constructed by AA-stacking the 2D Haldane honeycomb lattices. 
The orange and purple spheres denote the sublattice sites $A$ and $B$, 
respectively. In $xy$-plane, the black solid lines represent the NN hopping $%
t_{1}$, and the orange and purple arrows represent the NNN hopping $%
t_{2}\exp (\pm i\protect\phi )$. The orange and purple
lines represent interlayer hopping $t_{A}$ and $t_{B}$, respectively. 
(b) Brillouin zone of the 3D Chern insulator.}
\label{fig1}
\end{figure}

\emph{Lattice model}.---We study the tight-binding lattice 
model of a 3D photonic Chern insulator, which is
constructed by AA-stacking the 2D Haldane honeycomb 
lattices in the vertical direction \cite{PRL61-2015} and introducing the
interlayer hopping \cite{arxiv2106.02461} [Fig. \ref{fig1}(a)].
The on-site frequencies at sublattice sites $A$ and $B$ (orange and 
purple spheres) are $\omega_{A,B}$, respectively.
In $xy$-plane, the nearest-neighbour (NN) hopping (black lines)
is $t_{1}$, and the next-nearest-neighbour (NNN) 
hopping (orange and purple arrows) are $t_{2}e^{\pm i\phi }$.
In $z$ direction, the interlayer hopping for the
sublattice sites $A$ and $B$ are $t_{A}$ (orange lines) and $t_{B}$ (purple
lines), respectively. The lengths of the nearest-neighbor bonds in 
$xy$-plane and $z$ 
direction are $a_{0}$ and $h$, respectively. In the linear regime,
the Hamiltonian of this 3D photonic Chern insulator is
$H_{L} =  \sum_{i=0,1,2,3}d_{i}\sigma _{i}$,
where $d_{0} = \frac{\omega_{A}+\omega_{B}}{2} +(t_{A}+t_{B}) \cos(k_{z}h)
+2t_{2}\cos\phi \sum_{i=1,2,3} \cos(\mathbf{k} \cdot \mathbf{v}_{i})$,
$d_{1} = t_{1}\sum_{i=1,2,3} \cos(\mathbf{k} \cdot \mathbf{e}_{i})$,
$d_{2} = -t_{1}\sum_{i=1,2,3} \sin(\mathbf{k} \cdot \mathbf{e}_{i})$, and
$d_{3} = \frac{\omega_{A}-\omega_{B}}{2} +(t_{A}-t_{B}) \cos(k_{z}h)
-2t_{2}\sin\phi \sum_{i=1,2,3} \sin(\mathbf{k} \cdot \mathbf{v}_{i})$.
The two sets of vectors $\mathbf{e}_{1,2,3}$ and $\mathbf{v}_{1,2,3}$ are defined
for the NN hopping and NNN hopping in the horizontal plane, respectively.
Since we are interested in a 3D Chern insulator where the bulk bands are
characterized by a triad of Chern numbers $C=\left( C_{x},C_{y},C_{z}\right)
=\left( 0,0,1\right) $, in the following
we let $\omega_{A} = \omega_{B}$, $t_{A} = t_{B}>0$, and $\phi = \pi/2$.
Along the $KH$ and $K^{\prime}H^{\prime}$ lines in the Brillouin zone (BZ)
[Fig. \ref{fig1}(b)], this 3D photonic Chern insulator has linear dispersions
in the horizontal plane with $d_{1} = \mp v_{F} \delta k_{x}$,
$d_{2} = -v_{F} \delta k_{y}$, and
$d_{3} = \pm 3\sqrt{3}t_{2}$
according to $\mathbf{k}\cdot \mathbf{p}$ theory
 ("$-$" for $KH$ and "$+$" for $K^{\prime}H^{\prime}$).
Now the mass term is solely induced by the NNN hopping, which 
breaks the time-reversal symmetry. 
Here the group velocity $v_{F}$ is defined as
$v_{F}=\frac{\sqrt{3}}{2}t_{1}a$ with the transverse
lattice period $a = \sqrt{3} a_{0}$.
To study the dispersion in the vertical direction, we focus on the four
high-symmetry points: $K$, $K^{\prime}$, $H$, and $H^{\prime}$,  
since near these points the first order contributions are zero. From
$d_{0} = \frac{\omega_{A}+\omega_{B}}{2}
\pm (t_{A}+t_{B})(1 - \frac{h^{2}}{2}\delta k_{z}^{2})-3t_{2}\cos\phi$
(``$+$'' for $K$ and $K^{\prime}$; and ``$-$'' for $H$ and $H^{\prime}$),
this 3D photonic Chern insulator has quadratic dispersion in the vertical direction.
Specifically, it has anomalous GVDs at $K$ and $K^{\prime}$, and normal GVDs 
at $H$ and $H^{\prime}$.
Now the Hamiltonian $H_{L}$
with $d_{0,1,2,3}$ resembles the Hamiltonian in Eq. (\ref{Hamiltonian_1}),
except that the eigenfrequency $\omega$ is shifted by $\frac{\omega_{A}+\omega_{B}}{2}$.

Transforming the Hamiltonian $H_{L}$ into position space, we add a saturable nonlinear term
$H_{NL}=\text{diag} \left(N\left( \psi _{A}\right),N\left( \psi _{B}\right)\right)$
to $H_{L}$, where 
$N\left( \psi _{A,B}\right) =g\left\vert \psi
_{A,B}\right\vert ^{2}/\left( 1+\sigma \left\vert \psi _{A,B}\right\vert
^{2}\right) $ with the nonlinear parameter $g$, saturation coefficient $\sigma$,
and two pseudospin components $\psi _{A,B}$. 
Here we only focus on the self-focusing nonlinearity with $g>0$
(the case of self-defocusing nonlinearity with $g < 0$ can be studied similarly \cite{SM}). 
The existence of bright solitons requires
anomalous GVD in the vertical direction \cite{book-soliton}, 
which are fulfilled only at $K$ and $K^{\prime}$.
Thus, the whole Hamiltonian is 
\begin{equation}
H =  \sum_{i=0,1,2,3}\sigma _{i}d_{i},
\label{H_Chern}
\end{equation}
where 
\begin{eqnarray}
d_{0} &=& \frac{\omega_{A}+\omega_{B}}{2} + t_{A}+t_{B}
+ \frac{t_{A}+t_{B}}{2}h^{2}\partial_{z}^{2}
+\frac{N(\psi_{A})+N(\psi_{B})}{2}, \label{d0_Chern}\\
d_{1} &=& \pm iv_{F} \partial_{x}, \label{d1_Chern}\\
d_{2} &=& iv_{F} \partial_{y}, \label{d2_Chern}\\
d_{3} &=& \pm 3\sqrt{3}t_{2}+\frac{N(\psi_{A})-N(\psi_{B})}{2}. \label{d3_Chern}
\end{eqnarray}
Here, ``$\pm$'' correspond to $K$ and $K^{\prime}$, respectively.
Note that this Hamiltonian can also be derived directly from the coupled equations
in position space \cite{SM}.

\begin{figure}[tbp]
\includegraphics[width=8.3cm]{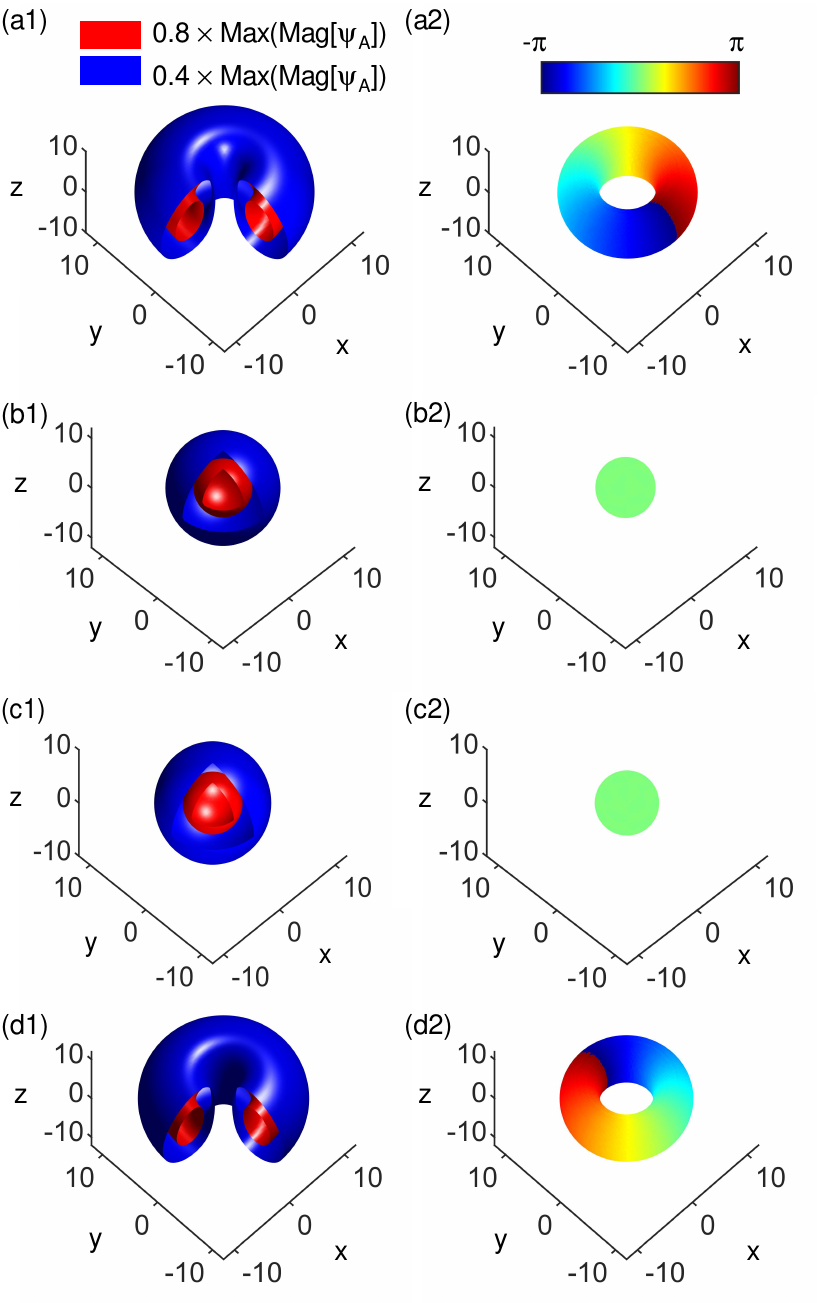}
\caption{(a1) Two different isosurfaces and (a2) the phase distribution on
the isosurface with $0.8*\text{Max}\left(\text{Mag}\left[\protect\psi_{A}%
\right]\right)$ of the pseudospin component $\protect\psi_{A}$ at $K$. 
(b1) Isosurfaces and (b2) the phase distribution of the pseudospin 
component $\protect\psi_{B}$ at $K$. 
(c1) Isosurfaces and (c2) the phase distribution of $\protect\psi_{A}$ at $K^{\prime}$. 
(d1) Isosurfaces and (d2) the phase distribution of $\protect\psi_{B}$ at $K^{\prime}$. 
The isosurfaces are plotted with $\omega=10$ and parts of the isosurfaces 
are removed for the sake of clarity.}
\label{fig2}
\end{figure}

Similar forms of Hamiltonian have been studied in free-space
Bose-Einstein condensates (BECs) with spin-orbit coupling (SOC), where the SOC terms are analogous
to the linear dispersions in $xy$ plane \cite{PRR2-013036}. 
However, in contrast to the externally imposed SOC, the linear dispersions 
are inherent in our lattice model.
We only study the fundamental solitons since the higher-order
solitons are usually unstable \cite{PRL116-214101}, and 
the parameters are $\omega _{A}=\omega _{B}=10$, 
$t_{1}=2/\sqrt{3}$, $t_{2}=1.05/3\sqrt{3}$, $%
t_{A}=t_{B}=0.5$, $a=1$, $h=1$, $g=1$, and $\sigma =10$. 
The topological solitons reside spectrally in the topological 
bandgap created by the linear bulk bands. Figs. \ref{fig2}(a1)-(b2) show the two pseudospin
components $\psi _{A,B}$ for the topological solitons at $K$ with $\omega =10$.
For the sake of clarity, parts of the isosurfaces are removed. 
From the isosurfaces [Figs. \ref{fig2}(a1) and (b1)],
in the horizontal plane the pseudospin component $\psi _{A}$ features a hump at a nonzero 
radius, and the component $\psi _{B}$ decreases monotonously in the radial direction.
This single hump behavior of our topological solitons is different to the soliton
profile in a Soler model \cite{PRA98-013827}, but they share the same
origin that nonlinearity induces a mass inversion and creates a topological domain 
wall in the bulk \cite{PRA98-013827,LPR13-1900223,arxiv1904.10312,
PRL111-243905,science368-856,SM}.  
While in $z$ direction, the topological solitons are self-localized because of the balance
between interlayer coupling and nonlinearity.
From the phase distributions on the isosurfaces [Figs. \ref{fig2}(a2) and (b2)], 
a vortex torus carrying a vorticity of $l_{A}=-1$ is formed for
the pseudospin component $\psi _{A}$, and the isosurface for 
$\psi _{B}$ is a sphere with a zero vorticity, namely $l_{B}=0$. The
vorticity (or topological charge) is defined as $l_{A,B}=\left( 1/2\pi
\right) \oint_{L}\nabla \left[ \arg \left( \psi _{A,B}\right) \right] \cdot d%
\vec{l}$. 
Thus, the topological solitons here are the
semi-vortex types \cite{NRP1-185}. Different to the semi-vortex BEC solitons
which are replaced by the Townes solitons in the absence of SOC \cite{PRL13-479}, 
our topological solitons vanish when the linear dispersion terms are mathematically
removed.

In Figs. \ref{fig2}(c1)-(d2) we show the topological solitons 
at high-symmetry point $K^{\prime}$ with $\omega = 10$.
For a 3D Chern insulator with time-reversal symmetry breaking (inversion symmetry is preserved), 
$d_{3}$ has an opposite sign at $K^{\prime}$ compared with the value of $d_{3}$ at $K$.
This leads to the equal Berry curvatures $\Omega$ at $K$ and $K^{\prime}$,
i.e.  $\Omega (\mathbf{k}) = \Omega (\mathbf{-k})$, which indicates a non-zero Chern
number $C_{z}$ \cite{PRL61-2015}.
Due to the same underlying topology, according to the Hamiltonian $H$ in Eqs. (\ref{H_Chern})-(\ref{d3_Chern}), 
if we make transformations $\psi_{A} \rightarrow -\psi_{B}$ and $\psi_{B} \rightarrow \psi_{A}$ 
to the equations at $K$, we can get the equations at $K^{\prime}$. From the figures, 
the component $\psi_{A}$ 
has a zero vorticity with $l_{A} = 0$, and the component $\psi_{B}$ carries a vorticity of $l_{B} = -1$.
Thus, the topological solitons at $K$ and $K^{\prime}$ both rotate 
clockwise with a phase difference of $\pi$. Note that for a 3D
valley-Hall insulator, the topological solitons at $K$ and $K^{\prime}$ rotate 
in opposite directions \cite{SM}.

\begin{figure}[tbp]
\includegraphics[width=8.3cm]{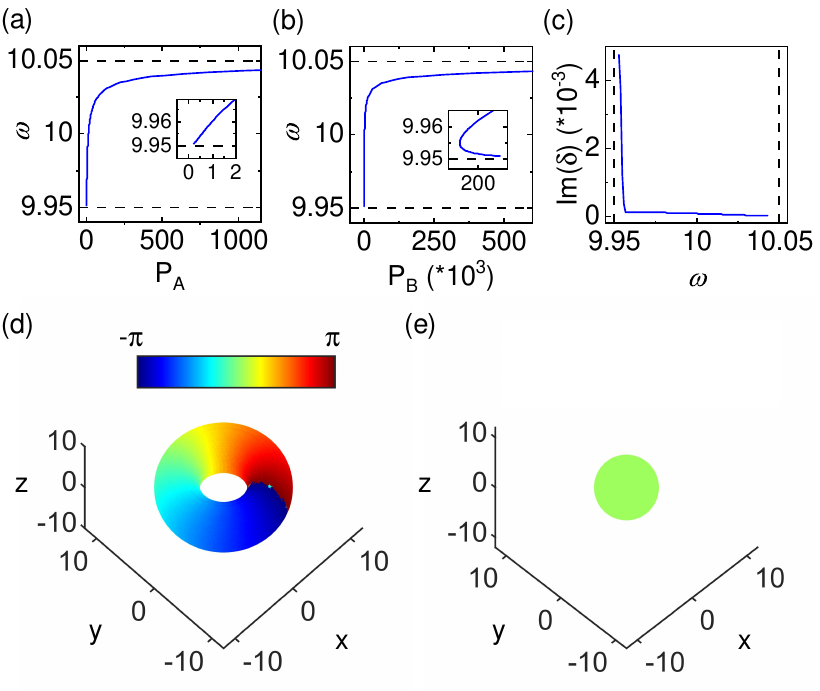}
\caption{(a)-(b) The power of the two pseudospin components $\protect\psi_{A,B}$. 
The insets are enlarged figures for $P_{A,B}$. 
(c) The growth rate $\text{Im} (\delta)$ of the topological solitons. 
The dashed lines in (a)-(c) denote the linear band edges.
(d)-(e) The perturbation eigenmode $\tilde{\varepsilon}_{A}$ 
and $\tilde{\varepsilon}_{B}$ at $\omega = 9.953$.}
\label{fig3}
\end{figure}

\emph{Existence and stability}.---In Figs. \ref{fig3}(a)-(b),
the frequency spectrum is plotted as a function of the 
powers $P_{A,B}$, 
which are defined as
$P_{A,B}=\int \left\vert \psi _{A,B}\left( \vec{r}\right) \right\vert^{2}d^{3}r$. 
We only show the plots for the topological solitons at $K$,
because the curves for the topological solitoins at $K^{\prime}$ can be obtained just by 
replacing $A (B)$ with $B (A)$. The dashed lines indicate the linear
band edges. The topological solitons bifurcate from the lower linear band edge
with a nonzero $P_{B}$, which implies that the topological solitons do not exist
below a certain power threshold. The family of topological solitons terminates
when the powers saturate. The power is
monotonic within most of the spectrum range. However, near the lower linear
band edge, 
we have $dP_{B}/d\omega<0$ [inset of Fig. \ref{fig3}(b))]. 
This negative slope is related to the stability of the topological solitons. 

We study the stability properties of the topological solitons using 
the linear stability analysis. 
The solution is sought at the frequency $\delta $ in the form of 
$\psi _{A,B}=\left( \phi _{A,B}+\varepsilon _{A,B}e^{-i\delta t}+\mu
_{A,B}^{\ast }e^{i\delta ^{\ast }t}\right) e^{-i\omega t}$,
where $\phi _{A,B}e^{-i\omega t}$ are the unperturbed soliton solution, $%
\varepsilon _{A,B}$ and $\mu _{A,B}$ are the perturbation eigenmodes. Note
that the perturbations may come from both the amplitudes and phases. For the
perturbation eigenmode with a certain vorticity $q$, the solution can be
written as%
\begin{eqnarray}
\left( 
\begin{array}{c}
\psi _{A} \\ 
\psi _{B}%
\end{array}%
\right) &=&\left[ \left( 
\begin{array}{c}
\tilde{\phi}_{A} \\ 
\tilde{\phi}_{B}%
\end{array}%
\right) +\left( 
\begin{array}{c}
\tilde{\varepsilon}_{A} \\ 
\tilde{\varepsilon}_{B}%
\end{array}%
\right) e^{-iq\varphi }e^{-i\delta t}+\left( 
\begin{array}{c}
\tilde{\mu}_{A}^{\ast } \\ 
\tilde{\mu}_{B}^{\ast }%
\end{array}%
\right) e^{iq\varphi }e^{i\delta ^{\ast }t}\right]  \notag \\
&&\left( 
\begin{array}{c}
e^{-i\varphi } \\ 
1%
\end{array}%
\right) e^{-i\omega t}.  \label{stability_eq_re}
\end{eqnarray}
Obviously, the topological solitons are linearly stable if $\delta $ is real. Whereas,
they are linearly unstable if the imaginary part of $\delta $, namely the
growth rate, is positive. From Fig. \ref{fig3}(c),
the topological solitons are linearly stable within most of the spectrum range. 
At a small regime near the lower linear band edge ($\omega < 9.957$), 
the topological solitons are linearly unstable because of the
emergence of nonzero imaginary part of $\delta$ via a Hopf bifurcation in
the $q=0$ spectrum at $\omega = 9.957$ [Figs. 3(d) and (e)]. Such instability is of the
exponential nature and can be predicted by the Vakhitov-Kolokolov criterion 
\cite{RQE16-783, PR142-103}, due to the fact
that the power $P_{B}$ dominates the total power and there is a negative
slope with $dP_{B}/d\omega<0$ near the lower linear band edge
[Fig. \ref{fig3}(b)]. This behavior is different to that of the topological solitons in 2D,
which are linearly stable near the lower linear band edge \cite{JPA50-495207}.

\begin{figure}[tbp]
\includegraphics[width=8.3cm]{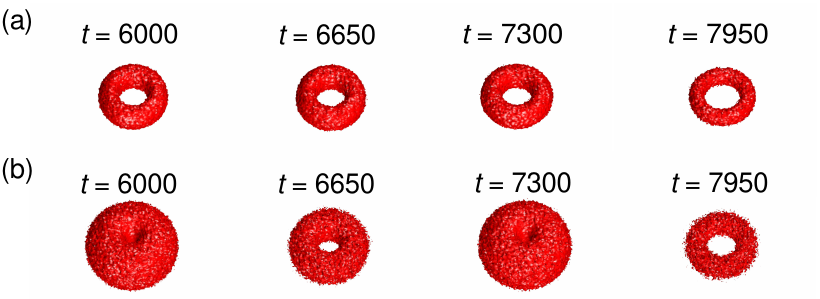}
\caption{(a) The isosurfaces with $0.8*\text{Max}\left(\text{Mag}\left[\protect\psi%
_{A} \left(t=0\right)\right]\right)$ of the pseudospin component $\protect%
\psi_{A} $ of the stable topological soliton with $\protect\omega=10$ at $K$. The four
subfigures from left to right correspond to $t = 6000$, $6650$, $7300$, and $%
7950$, respectively. (b) The isosurfaces with $1.0*\text{Max}\left(%
\text{Mag}\left[\protect\psi_{A} \left(t=0\right)\right]\right)$ of 
$\protect\psi_{A}$ of the unstable topological soliton with 
$\protect\omega=9.953$.}
\label{fig4}
\end{figure}

\emph{Dynamics}.---We add $\pm 10 \%$ noises with uniform distributions 
to the topological solitons at $K$ and study their temporal evolution. In
Figs. \ref{fig4}(a)-(b), we show the isosurfaces of the pseudospin component $%
\psi_{A}$ at different times with $\omega=10$ and $\omega=9.953$,
respectively. For the stable topological soliton with $\omega=10$, although
noises are imposed, the soliton is always self-sustained in all 3D and the radius of the
torus tube is invariant [Fig. \ref{fig4}(a)]. For the unstable topological
soliton with $\omega=9.953$, it exhibits a breathing structure [Fig. \ref{fig4}(b)]. 
The radius of the torus tube and the magnitude of
the soliton oscillate along with the temporal evolution. Since the growth rates $%
\text{Im}\left(\delta\right)$ are in the order of $10^{-3}$, the topological
solitons near the lower linear band edge are weakly unstable. Thus, our topological
solitons in 3D should be observable in the whole spectrum
range. Furthermore, although the topological solitons are semi-vortex
solitons where the pseudospin component $\psi_{A}$ have a nonzero vorticity,
they are only disturbed by the radially symmetric perturbations with $q = 0$
and radial symmetry breaking is not observed. This behavior agrees with the
result from the linear stability analysis.

\emph{Conclusion}.--- We find self-localized topological states 
(i.e. topological solitons) in a 3D nonlinear photonic Chern insulator.
The topological solitons at high-symmetry points $K$ and $K^{\prime}$ 
rotate in the same direction, as a manifestation of the topology of the linear
host lattice. Specifically, these solitons are stable over a broad frequency range. 
Because of these features, it is feasible to observe the topological solitons
experimentally. Considering
that both time-reversal symmetry breaking and nonlinearity can be
implemented in electrical circuit lattices \cite{PRL122-247702,
nelectron1-178} and 3D circuit lattices are
readily available \cite{NSR8-nwaa192}, we propose a realistic circuit
implementation to observe the topological solitons \cite{SM}.
Our work establishes how the interplay between topology and nonlinearity
leads to a new type of solitons, and can be extended to other 
3D topological systems.

\bigskip

\begin{acknowledgments}
R.L. was sponsored by the National Natural Science Foundation of China 
(NSFC) under Grant No. 12104353. P.L. was sponsored by the National Natural 
Science Foundation of China (NSFC) (11805141), the Applied Basic Research 
Program of Shanxi Province (201901D211424), and the Scientific and Technological 
Innovation Programs of Higher Education Institutions in Shanxi (STIP) (2021L401). 
Y.L. was sponsored by the National Natural Science Foundation of China (NSFC) 
under Grant No. 61871309 and the 111 Project. The numerical calculations in this paper 
was supported by High-Performance Computing Platform of Xidian University. 
Part of the computation is supported by Shenzhen Bkunyun Cloud Computing Co., Ltd.
\end{acknowledgments}

\end{document}